# High Mobility WSe$_2$ *p*- and *n*-Type Field Effect Transistors Contacted by Highly Doped Graphene for Low-Resistance Contacts

Hsun-Jen Chuang[1], Xuebin Tan[2], Nirmal Jeevi Ghimire[3,4], Meeghage Madusanka Perera[1,] Bhim Chamlagain[1], Mark Ming-Cheng Cheng[2], Jiaqiang Yan[3,4], David Mandrus[3,4], David Tománek[5] and Zhixian Zhou[1, a)]

[1]Department of Physics and Astronomy, Wayne State University, Detroit, MI 48201
[2]Department of Electrical and Computer Engineering, Wayne State University, Detroit, Michigan 48202
[3]Deptment of Materials Science and Engineering, The University of Tennessee, Knoxville, TN 37996
[4]Materials Science and Technology Division, Oak Ridge National Laboratory, Oak Ridge, TN 37831
[5]Physics and Astronomy Department, Michigan State University, East Lansing, MI 48824

a) Author to whom correspondence should be addressed, electronic mail: zxzhou@wayne.edu

Abstract

We report the fabrication of both n-type and p-type WSe$_2$ field effect transistors with hexagonal boron nitride passivated channels and ionic-liquid (IL)-gated graphene contacts. Our transport measurements reveal intrinsic channel properties including a metal-insulator transition at a characteristic conductivity close to the quantum conductance e$^2$/h, a high ON/OFF ratio of >10$^7$ at 170 K, and large electron and hole mobility of μ ≈ 200 cm$^2$V$^{-1}$s$^{-1}$ at 160 K. Decreasing the temperature to 77 K increases mobility of electrons to ≈330 cm$^2$V$^{-1}$s$^{-1}$ and that of holes to ≈270 cm$^2$V$^{-1}$s$^{-1}$. We attribute our ability to observe the intrinsic, phonon limited conduction in both the electron and hole channels to the drastic reduction of the Schottky barriers between the channel and the graphene contact electrodes using IL gating. We elucidate this process by studying a Schottky diode consisting of a single graphene/WSe$_2$ Schottky junction. Our results indicate the possibility to utilize chemically or electrostatically highly doped graphene for versatile, flexible and transparent low-resistance Ohmic contacts to a wide range of quasi-2D semiconductors.

Layered transition metal dichalcogenides (TMDs) have recently emerged as promising materials for flexible electronics and optoelectronics applications. These systems have demonstrated many "graphene like" properties including a relatively high carrier mobility, mechanical flexibility, chemical and thermal stability, and moreover offer the significant advantage of a substantial band gap.[1] Field-effect transistors (FETs) with atomically thin TMD channels are immune to short channel effects.[2] In addition, pristine surfaces of two-dimensional (2D) TMDs are free of dangling bonds, which reduce surface roughness scattering and interface traps. Atomic layers of $MoS_2$ are probably the most extensively studied among the layered TMDs due to the availability of large natural molybdenite crystals from mining sources.[3] In addition to $MoS_2$, several other semiconducting TMDs such as $MoSe_2$, $WS_2$ and $WSe_2$ with different band structures and charge neutrality levels may offer additional distinct properties.[1, 4] However, the number of studies on TMDs other than $MoS_2$ is still small.[5-14] Among these studies, back-gated $WSe_2$ monolayer FETs with surface doping have already demonstrated a high field-effect mobility[8] reaching ~140 $cm^2V^{-1}s^{-1}$, which is substantially higher than most of the reported room-temperature mobility values for $MoS_2$.[15-19] A high intrinsic hole mobility of up to 500 $cm^2V^{-1}s^{-1}$ was also observed in bulk $WSe_2$ FETs.[11] Furthermore, $WSe_2$ is more resistant to oxidation in humid environments than $MoS_2$.[10, 20]

A major challenge for developing $WSe_2$-based electronic devices is that $WSe_2$ tends to form a substantial Schottky barrier (SB) with most metals commonly used for making electrical contacts.[8, 13] This is a strong disadvantage, since low-resistance Ohmic contacts are needed for exploring i) intrinsic transport properties of the channel material, and ii) performance limits of realistic devices. There are typically two approaches to achieve low resistance contacts between a semiconductor and a metal: (a) thinning the SB width by degenerately doping the contact regions and (b) lowering the SB height by selecting contact materials with an extremely high (for *p*-type semiconductors) or low (for *n*-type semiconductors) work function. Low-resistance contacts have been achieved by surface doping of $WSe_2$ (e.g. with $NO_2$ and K) in order to reduce the SB thickness, or by using low work function contact metals such as Indium in order to lower the height of the SB to the conduction band.[8, 10, 12] However, $NO_2$ and K doping is not stable in air. Indium adheres poorly to the substrate and is thermally unstable due to its low melting point of 156ºC.[10] Previously, we have reported significant improvement of electrical contacts in few-layer $MoS_2$ devices by drastically reducing the SB thickness using an ionic liquid (IL) gate.[21] However, the improved charge injection efficiency using this method is still fundamentally limited by the height of the Schottky barrier. For example, the contact resistance of our IL-gated $MoS_2$ is still significantly higher for the hole channel than for the electron channel due to the relatively large SB height for the valence band. As mentioned earlier, the Schottky barrier height may be reduced by selecting metal electrodes with a low work function (for n-type semiconductors) or a high work function (for p-type semiconductors). Yet it has proven to be extremely challenging to find metals with a proper high or low work function that also exhibit a high conductivity and a high chemical, thermal and electrical stability. Furthermore, the expected benefit of the proper work function for lowering the SB may be drastically reduced by Fermi level pinning.[2, 22] In particular, a recent theoretical study shows that partial Fermi level pinning is present at the metal/TMD contacts for a variety of metals with the work functions spanning a wide range.[23]

In this work, we use graphene as electrode material with a tunable work function to overcome the above limitations of metal electrodes to contact few-layer $WSe_2$ FETs. For one, graphene is mechanically strong, flexible and thermally stable, which is desirable for flexible electronics applications. Even more importantly, the work function of graphene can also be tuned by chemical or electrostatic doping to minimize the SB height at the graphene/semiconductor interface.[24, 25] By using the extremely large electric double layer (EDL) capacitance of an IL gate, the work function of graphene at the graphene/$WSe_2$ interface can be modulated within an enormously large range. As a result, we have formed for the first time, in a single device structure, $WSe_2$-based FETs of both *n*- and *p*-type that display low-resistance contacts and high carrier mobility. Even though devices based on graphene-TMD heterostructures have been reported previously,[26-28] the operation principle of our $WSe_2$ FETs with lateral graphene contacts is qualitatively different from the previously reported graphene-based field-effect tunneling transistors (FETTs) and vertical field-effect transistors (VFETs). Whereas transport in FETTs and VFETs based on graphene/TMD heterostructures is modulated by the vertical transport barrier at the graphene/TMD junctions, the graphene/TMD junctions in our devices serve as optimum drain and source contacts to lateral WSe$_2$ FETs. We also emphasize that IL gating is crucial to achieving a sufficiently large modulation of the graphene work function that is needed for making low resistance contacts to both the valence and conduction bands of WSe$_2$. Although back-gated MoS$_2$ FETs with graphene contacts have also been reported recently, achieving such a large modulation of the work function is not possible with a back gate alone.[29-32] In addition to FET operations, we have also demonstrated Schottky diode operations in our WSe$_2$ devices by applying asymmetric IL gate voltages to the two graphene/WSe$_2$ contacts.[33] Our Schottky diode device displays a significant rectifying behavior and a diode ideality factor of ≈1.3, signifying a high quality of the graphene/WSe$_2$ interface. Variable temperature electrical measurements performed on WSe$_2$ FETs with low-resistance graphene contacts reveal that, as the temperature decreases from 160 K to 77 K, the extrinsic field-effect mobility increases from ≈196 cm$^2$V$^{-1}$s$^{-1}$ to ≈330 cm$^2$V$^{-1}$s$^{-1}$ for the electron channel and from ≈204 cm$^2$V$^{-1}$s$^{-1}$ to ≈270 cm$^2$V$^{-1}$s$^{-1}$ for the hole channel. These mobility values are comparable to the highest electron and hole mobilities reported on thin film WSe$_2$ FETs as well as bulk WSe$_2$,[8, 10, 12, 34] indicating that our graphene-contacted few-layer WSe$_2$ devices approach the intrinsic phonon-limited mobility for both electrons and holes. The most significant finding of our study is that highly "doped" graphene is an excellent contact electrode material for high-performance *p*- and *n*-type TMD FETs. The different functionalities of highly *n*-doped and highly *p*-doped graphene electrodes offer a promising possibility to fabricate complementary digital circuits on a single TMD thin film.

To fabricate graphene contacted WSe$_2$ FETs, atomically thin WSe$_2$ flakes were produced from a bulk crystal by a mechanical cleavage method and subsequently transferred onto degenerately doped silicon substrates covered with a 290 nm thick thermal oxide layer.[21, 35-37] An optical microscope was used to identify thin flakes, which were further characterized by non-contact-mode atomic force microscopy (AFM). In the present study, we focus on few-layer samples with 4-12 layers corresponding to 3-8 nm thickness. Samples containing only few layers can be produced more easily and sustain larger drive currents than WSe$_2$ monolayers, while at the same time maintaining a relatively large ON/OFF ratio and a small *c*-axis interlayer resistance in comparison to thicker samples.[31, 38, 39] Next, we deposited a thin hexagonal boron nitride (h-BN) crystal (10–50 nm thick; mechanically exfoliated from commercially available h-BN crystals) onto a few layer $WSe_2$ flake to cover its middle section. While performing the deposition using a home-built precision transfer stage, we made sure that the two ends of the $WSe_2$ remained exposed to form electrical contacts. Passivating the $WSe_2$ channel by h-BN enables us to tune separately the SB height at the graphene drain/source contacts using the IL gate and the chemical potential of the channel using the Si back gate. We chose h-BN as the $WSe_2$ channel passivation layer, because of its atomically smooth surface that is chemically inert and relatively free of charged impurities and charge traps.[40-42] To form drain and source contacts, we transferred patterned CVD grown monolayer graphene on top of the h-BN covered $WSe_2$ flake. Subsequently, we fabricated metal electrodes, consisting of 5 nm of Ti covered by 50 nm of Au, as electrical contacts to the graphene electrodes using standard electron beam lithography (EBL) and electron beam deposition.[43] During the same fabrication step, a large metal electrode was fabricated to serve as an IL gate electrode. Finally, we removed selectively graphene on top of the h-BN covered $WSe_2$ channel by EBL patterning and oxygen plasma etching. Then, a small droplet of the DEME-TFSI IL (Sigma Aldrich 727679) was carefully deposited onto the devices using a micromanipulator under an optical microscope, covering the $WSe_2$ devices and gate electrodes.[21]

Figure 1a shows a schematic illustration and Figure 1b shows a micrograph of a typical $WSe_2$ FET device with graphene contacts. Graphene is located inside the rectangular box outlined by the white dashed lines. Figure 1c shows an AFM image of the region boxed by the black solid lines in Figure 1b, consisting of the h-BH covered channel, graphene drain and source contacts, and a part of the Au leads contacting the graphene electrodes. A large part of the graphene/$WSe_2$ drain and source contacts is clean and free from bubbles, with a relatively low RMS roughness of a few tenths of a nanometer, indicating a high quality of the graphene/$WSe_2$ interface. The few wrinkles observed in the graphene/$WSe_2$ contact areas in Figure 1c are not expected to have a significant impact on the overall contact quality. This is indeed verified by the near-ideal Schottky diode behavior in a graphene contacted $WSe_2$ device, which is dominated by a single graphene/$WSe_2$ junction as discussed in detail below. Electrical properties of the devices were measured by a Keithley 4200 semiconductor parameter analyzer in a Lakeshore Cryogenic probe station after dehydrating the IL under high vacuum (~$1\times10^{-6}$ Torr) for 48 hours. This removed any remaining moisture thoroughly, which turned out to be important for preventing the formation of chemically reactive protons and hydroxyls through the electrolysis of water.[21, 44] As shown schematically in Figure 1a, negative ions in the IL accumulate near the gate electrode and positive ions accumulate near the $WSe_2$ channel when a positive voltage $V_{ILg}$ is applied to the IL gate electrode near the device channel. This scenario reverses when a negative voltage is applied to the gate. In both cases, electric double layers form at the interface between the IL and the solid surfaces.[21, 45] We have not measured electron and hole mobilities at room temperature to avoid the possibility of chemical reactions involving the IL at high voltages. Since the electrochemical stability of the IL increases with decreasing temperature, larger IL gate voltages can be applied without causing chemical reactions at lower temperatures. To minimize the possibility of chemical reactions between the IL and graphene contacts, electrical transport measurements were carried out between 77 and 180 K, after the devices had been quickly cooled from 230 K to below 170 K at fixed IL voltage. Below the freezing point of the IL at ≈200 K, the carrier density induced by the presence of positive or negative ions, which preferentially enriched the vicinity of $WSe_2$, remained practically constant.[21] Measuring the electrical property of IL-gated devices below the freezing point of the IL also eliminates possible coupling between the Si back and IL gate.[21, 37]

We have measured several $WSe_2$ devices with IL-gated graphene drain and source contacts. Among these, two devices had h-BN channel passivation, one had its channel covered by a 50 nm thick $Al_2O_3$ layer, and a few other devices had bare channels (uncovered). While qualitatively consistent results were observed in all devices, the two h-BN covered $WSe_2$ devices showed the best performance, which can be attributed to the excellent interface quality between $WSe_2$ and h-BN. Slightly inferior results obtained using the $Al_2O_3$ passivated $WSe_2$ device and an unpassivated, bare $WSe_2$ device are discussed in the Supplementary Information.

Figure 2a shows the transfer characteristics of a 6 nm thick h-BN passivated $WSe_2$ sample with a channel length of 4.8 μm, measured at room temperature at $V_{ds}$ = 0.1 V before the IL has been deposited.

The device shows clear ambipolar behavior with relatively low drain-source current for both the hole and electron channels, indicating significant SBs in both the valence and conduction band regions of $WSe_2$. In this case, both electron and hole injection is accomplished either via thermal excitation across the SB or tunneling at the band edge, or by thermally assisted tunneling, which combines the first two mechanisms. Consequently, electrons (holes) are preferably injected into the conduction (valence) band as the back-gate induced band bending reduces the SB thickness, and as the back-gate modulation of the graphene carrier density—which shifts its Fermi level—lowers the SB height at high positive (negative) gate voltages. This SB modulation is schematically illustrated in the inset of Figure 2a. It is imperative to further reduce the height of SBs at the graphene/$WSe_2$ contacts, since the presence of a significant contact resistance associated with a large SB height prevents the exploration of the intrinsic performance limits of $WSe_2$ as a FET channel material. To minimize the SB height and thus the contact resistance, we applied large positive (negative) IL gate voltages to the graphene contacts to optimize the electron (hole) injection to $WSe_2$. Figure 2b shows the transfer characteristics of the device measured at 170 K and in the Si back gate configuration for different IL gate voltages applied during cooling down from 230 K to 170 K. Since the $WSe_2$ channel is protected from direct contact with the IL by a h-BN crystal, the difference in the $I_{ds}$ $-V_{gs}$ characteristics for different $V_{ILg}$ arises chiefly from the IL-gate tuning of the graphene/$WSe_2$ contacts. Without applying a voltage to the IL gate, even though the graphene contacts are already covered by IL, the device shows a similar ambipolar behavior with comparable electron and hole currents as prior to the deposition of the IL droplet. As the IL gate voltage increases to $V_{ILg}$ = 6 V, the ON-current for the electron channel, measured at $V_{bg}$ = 70 V, increases by an order of magnitude to 3 µA/µm, while the hole current measured at $V_{bg}$ = -70 V decreases by over three orders of magnitude from $10^{-2}$ µA/µm to below $10^{-5}$ µA/µm. The ON/OFF ratio of the device exceeds $10^7$ for the electron channel, while that for the hole channel is less than $10^2$ when cooled down at $V_{ILg}$ = 6 V. This behavior is associated with opposite trends in the changing height of the SB to the conduction and valence bands, which is caused by the IL gate modulation of the carrier density that changes the Fermi level in the graphene drain and source contacts. As the positive IL gate voltage increases, the Fermi level in graphene shifts upward, thus decreasing (increasing) the SB height for the conduction (valence) band. Moreover, the threshold voltage decreases with increasing IL gate voltage as shown in the inset of Figure 2b, which can also be attributed to the reduction of SB height for the conduction band, because a higher $V_{bg}$ is needed to overcome a larger SB height to inject electrons into the conduction band, chiefly via band bending. The hysteresis in the room temperature transfer characteristics, shown in Figure 2a, likely originates from charge trapping in the $SiO_2$ substrate. This is corroborated by the observation that the hysteresis diminishes below 170 K, where the charge trapping is suppressed (see Figure 2b).[46-48]

The output characteristics of the $WSe_2$ device after being cooled down to T=170 K is shown in Figure 2d for zero applied IL gate voltage and in Figure 2e for $V_{ILg}$= 6 V. As shown in Figure 2d, the drain current displays strongly non-linear (upward turning) $I_{ds}$–$V_{ds}$ behavior in the case of no applied IL gate voltage, suggesting the presence of a significant SB at the contacts. In sharp contrast to this, the $I_{ds}$ − $V_{ds}$ characteristics of the same device, when cooled down at $V_{ILg}$= 6 V, is linear at all back gate voltages according to Figure 2e. Furthermore, the resistance of the device calculated from the slope of the $I_{ds}$ - $V_{ds}$ characteristics, measured in the low-bias region at $V_{bg}$ = 40 V with no applied IL voltage, is over 2 orders of magnitude larger than for $V_{ILg}$ = 6 V. Also, the ON-current of 0.27 µA/µm, measured at $V_{ds}$= 1V and $V_{bg}$=40 V with no applied IL gate voltage, is about 67 times smaller than the ON-current value of 18 µA/µm for $V_{ILg}$=6 V. Both observations provide further evidence that a large positive value of $V_{ILg}$ reduces the contact resistance for the electron channel, most likely by primarily lowering the SB height. As depicted schematically in the insets of Figures 2d and 2e, the Fermi level in graphene is close to the middle of the $WSe_2$ band gap with no applied IL voltage, giving rise to a significant SB. When $V_{ILg}$ = 6V is applied to the IL gate, the Fermi level of graphene lines up with (or close to) the conduction band edge, forming a very low SB. Note that the ON-current at $V_{ILg}$ = 6 V, $V_{bg}$ = 40 V and $V_{ds}$= 1 V shows no sign of saturation, indicating that higher ON-currents are possible, as we expand the voltage range in the Supplementary Information. Figures 2c and 2f show the transfer and output characteristics of the same devices at $V_{ILg}$ = -7 V after cooling down from 230 K to 170 K. When the graphene/$WSe_2$ contacts are gated by a large negative voltage, the device exhibits clear *p*-type conduction with a high current ON/OFF ratio exceeding $10^7$. In this case, the SB for the valence band carrier injection is significantly reduced, since the Fermi level of the graphene electrode has been lowered to line up with (or close to) the valence band edge, as depicted in the inset of Figure 2f. As shown in the main panel of Figure 2f, the $I_{ds}$-$V_{ds}$ characteristics measured in the hole channel are also linear and exhibit similar drain-source currents as previously observed for the electron channel at $V_{ILg}$ = 6 V. Our findings indicate that low resistance contacts were achieved also for the valence band. It is worth pointing out that the ideal scenario of zero SB height, as schematically illustrated in the insets of figures 2e and 2f, is unnecessary for achieving low-resistance contacts in our WSe$_2$ samples. It is conceivable that a small SB still exists in our graphene-contacted WSe$_2$ devices, where the contact resistance is drastically reduced by the combined effect of SB height reduction and SB width narrowing due to IL gating. The graphene/WSe$_2$ contact resistance was estimated from the output characteristics of a short channel (L≈200 nm) graphene-contacted WSe$_2$ device at high $V_{bg}$ and high $V_{ILg}$ (see Supplementary Information for details). The extracted contact resistance is less than 2kΩ·µm, which is substantially lower than the reported contact resistance of $7\times10^5$ Ω.µm for Ti electrodes and $6.5\times10^3$ Ω.µm for Ag electrodes in multilayer WSe$_2$ FETs.[10]

To further characterize the graphene/WSe$_2$ interface quality and to shed additional light on the electric field modulation of graphene/WSe$_2$ junctions, we fabricated a WSe$_2$ device that is dominated by a single graphene/WSe$_2$ SB. A Schottky diode behavior in this device was achieved by applying an asymmetric IL gate voltage to create one high-resistance SB contact and one low-resistance contact to the two electrodes.[33] Figure 3a shows typical Schottky diode characteristics of an *n*-type graphene/WSe$_2$ SB junction measured at back gate voltages $V_{bg}$ = 20 V and 30 V. Moderate positive back-gate voltages were chosen to turn the electron channel on so that the resistance of the entire device was dominated by the high-resistance graphene/WSe$_2$ SB junction. A diode ideality factor of ≈ 1.3 was obtained in the forward region at low bias voltages. The closeness to the ideality factor of 1.0 obtained for this device is comparable to that of optimized graphene/Si contacts,[25] further validating the high interface quality of our graphene/WSe$_2$ contacts. Realization of a SB diode simplifies the procedure to extract the SB height from the temperature dependence of the diode characteristics, without the complications arising from back-to-back SB contacts in most TMD FET devices. WSe$_2$ in contact with the graphene drain and source electrodes is likely electrostatically doped by the back gate voltage, which has been applied to turn on the channel, and possibly also by the IL gate voltage in view of the incomplete screening by graphene. In this case, the current injection through the graphene/WSe$_2$ SB junction is expected to be dominated by thermally assisted tunneling rather than thermionic emission. Therefore, the commonly used thermal activation method is expected to underestimate the SB height in our SB diode. To obtain a better estimate of the SB height at the graphene/WSe$_2$ junction, we have adopted a model that considers the thermally-assisted tunneling current to highly doped semiconductors through a SB.[49, 50] Figure 3b shows forward *I-V* characteristics in the exponential region, which appears linear in the semi-logarithmic plot, measured at temperatures between 77 and 160 K within one thermal cycle in order to avoid possible thermal hysteresis effects. Our data are well represented by the equation $I_f = I_s \exp(eV_f/\Phi_0)$, where $I_s$ is the saturation current at zero forward voltage $V_f = 0$ V and $\Phi_0$ is a characteristic energy, both of which are strongly temperature dependent. Furthermore, the saturation current $I_s$ also depends exponentially on $1/\Phi_0$. Figure 3c shows the semi-logarithmic plot of $I_s$ as a function of $1/\Phi_0$ using the $I_s$ and $\Phi_0$ values obtained from the fits of our data at different temperatures for $V_{bg} = 20$ V, shown in Figure 3b, and $V_{bg} = 30$ V. Fitting the data for the saturation current $I_s$ in Figure 3c by the function $I_s \sim \exp(\Phi_b/\Phi_0)$, we obtain $\Phi_b$=0.44 eV as a SB height of the graphene/WSe$_2$ junction at $V_{bg}$= 20 V and the lower value $\Phi_b$=0.31 eV at $V_{bg}$=30 V. The SB height obtained here is consistent with the difference between the graphene work function and the conduction band minimum in WSe$_2$.[4] The reduction of SB height at increasing back gate voltage can be attributed to the back gate tuning of the graphene work function, since the relatively low carrier density in the thin WSe$_2$ layer is not sufficient to screen the back gate electric field. In the absence of Fermi level pinning and electric field screening, the maximum tunability of the graphene work function is estimated to be of the order of 0.1 eV near the charge neutrality point of a graphene monolayer for a back-gate voltage range of 10 V in our devices (see the Supplementary Information). The work function of the graphene contacts is also rather insensitive to the back gate voltage, and is chiefly determined by the IL gate at high $V_{ILg}$ values due to the extremely high carrier density in graphene and the much larger IL gate capacitance. The fact that the graphene work function changes by the theoretical limit as the back gate voltage changes by 10 V indicates that Fermi level pinning is nearly absent in our graphene/WSe$_2$ junctions and that graphene is close to its charge neutrality point.

The low-resistance contacts fabricated in our study by highly doping graphene at the graphene/WSe$_2$ junctions using large IL-gate voltages enable us to investigate for the first time the intrinsic properties of both the hole and electron channels in TMDs on the same device. Figures 4a and 4b show the electron and hole channel conductivity of the same device, characterized in Figure 2, as a function of back gate voltage between 77 and 160 K. The conductivity is defined by $\sigma = I_{ds}/V_{ds} \times L/W$, where $L$ is the length and $W$ the width of the h-BN passivated WSe$_2$ channel. The conductivity increases with decreasing temperature for electron channels at high positive and for hole channels at high negative back gate voltages. Interestingly, with increasing hole density, we observe a crossover from an insulating regime, where the conductivity increases with increasing temperature, to a metallic regime, where the conductivity decreases with temperature. This crossover occurs at the critical conductivity of $e^2/h$, consistent with the observed metal-insulator transition (MIT) in monolayer, bilayer and multilayer MoS$_2$ as well as theoretical expectations for 2D semiconductors.[16, 19, 51] However, the MIT in our WSe$_2$ device occurs at a much lower carrier density of $\approx 1 \times 10^{12}$ cm$^2$ than that in MoS$_2$ ($\approx 1 \times 10^{13}$ cm$^2$), which may be attributed to the lower level of impurity states inside the band gap of our h-BN passivated WSe$_2$ devices than in previously reported MoS$_2$ devices.[16] Here, the critical carrier density for the MIT is determined by $C_{bg} \times (V_{MIT} - V_{th})/e$, where $V_{MIT}$ is the back gate voltage of the MIT and $V_{th}$ is the threshold voltage. Unlike the hole channel, this MIT is considerably suppressed for the electron channel, as seen in Figure 2a. As our WSe$_2$ crystals are likely slightly *p*-doped, most impurity states in the band gap are expected to be closer to the valence band edge. As a result, any disorder should affect the conduction band less than the valence band. Furthermore, the threshold voltage for the conduction band is close to $V_{bg} = 0$ V, whereas it is close to -20 V for the valence band, confirming further that there are more impurity states in the gap near the valence band edge than the conduction band edge. As the chemical potential of the WSe$_2$ channel is tuned toward the valence band by the back gate, it crosses localized band tail states before reaching the valence band, leading to a non-negligible threshold voltage. On the other hand, the substantially lower density of impurity states near the conduction band edge allows the electron channel to be turned on much more quickly by a gate voltage, leading to a significantly reduced threshold voltage.[52] Given the small hysteresis in the transfer characteristics of the device, seen in Figures 2b and 2c, our observation of a MIT in the hole channel is unlikely a hysteretic effect. Indeed, the conductivity curves for opposite directions of gate voltage sweeps, taken from the dual-sweep transfer characteristics at 77 K < T < 160 K, show nearly identical behavior, as seen in Figure S5 in the Supplementary Information. This observation further rules out the possibility of a hysteresis influencing the presence and absence of a MIT for hole and electron channels.[16] Figure 4c shows the temperature dependence of the mobility for both the hole and electron channels, extracted from the linear region of the conductivity curves in the metallic state, corresponding to -45 V < $V_{bg}$ < -35 V for the hole channel and 10 V < $V_{bg}$ < 20 V for the electron channel, using the expression for field-effect mobility $\mu=(1/C_{bg})\times(d\sigma/dV_{bg})$. As the temperature decreases from 160 K to 77 K, the mobility in the electron channel increases from 196 $cm^2V^{-1}s^{-1}$ to about 330 $cm^2V^{-1}s^{-1}$ and that of the hole channel increases similarly from 204 $cm^2V^{-1}s^{-1}$ to about 270 $cm^2V^{-1}s^{-1}$. The mobility values observed here are comparable to the highest mobility values reported in thin film $WSe_2$ transistors.[8, 10, 12] The increase of the mobility with decreasing temperature along with the high absolute values of the mobility strongly suggests that we are observing the intrinsic channel properties that are dominated by phonon scattering. It is worth pointing out that also the channel passivation with ultra-flat and ultraclean h-BN crystals is critical to achieving high mobilities and low threshold voltages by minimizing the presence of interface trap states. As we discuss in the Supplementary Information, substantially lower mobilities and higher threshold voltages were observed in $WSe_2$ devices, where the channel was passivated by a 50 nm thick $Al_2O_3$ layer.

In conclusion, we have studied the use of graphene as a work-function-tunable electrode material for few-nanometer thick $WSe_2$ FETs. We have shown that the contact resistance at the graphene/$WSe_2$ drain and source contacts can be drastically reduced by lowering the Schottky barrier height by IL gating. Our Schottky diode fabricated by applying asymmetric IL gate voltages to the source and drain contacts exhibits near-ideal SB diode behavior, further corroborating that the graphene/WSe$_2$ interface is of high quality. Realization of low-resistance contacts along with the channel passivation using ultra-clean h-BN, has enabled us to observe intrinsic charge transport properties of the WSe$_2$ channel. The approach demonstrated here can be applied to a wide range of layered two-dimensional semiconductors in addition to TMDs. We use IL gating as a highly effective and versatile method to demonstrate one way to achieve low-resistance contacts by significantly reducing the SB height at the graphene/WSe$_2$ drain and source electrodes. Methods to achieve high electron- and hole-doping of graphene that yield more permanent and air-stable contacts will be explored in the future.

**Supporting Information**

Supporting information contains details of the fabrication process for all graphene-contacted WSe$_2$ FET devices, additional transport data on graphene-contacted WSe$_2$ devices and graphene FETs, estimation of graphene/WSe$_2$ contact resistance, and discussion of graphene work function change by gating.

*Conflicts of interest:* The authors declare no competing financial interest.

*Acknowledgement*

This work was supported by NSF (ECCS-1128297 and DMR-1308436 ). XT and MMCC were supported by NSF CAREER Award (1055932) MRI Award (1229635) and Wayne State University. NJG, JY and DM were supported by Materials and Engineering Division, Office of Basic Energy Sciences, U.S. Department of Energy. DT was supported by the National Science Foundation Cooperative Agreement #EEC-0832785, titled "NSEC: Center for High-rate Nanomanufacturing". We would also like to thank Dr. Ir. Niko Tombros for the technical advice on building the microcrystal transfer apparatus used in this work.

**Figure Captions:**

**Figure 1.** (a) Schematic illustration of the structure and working principle of a WSe$_2$ FET with ionic-liquid-gated graphene contacts. (b) Optical micrograph of a typical WSe$_2$ FET with graphene electrodes before applying the ionic liquid. The contours of the graphene are marked by white dashed lines. The scale bar is 10 µm. (c) AFM image of the device region enclosed in the black box in (b). The scale bar is 1 µm.

**Figure 2.** Transfer and output characteristics of a 6 nm thick WSe$_2$ FET device with graphene contacts and a 4.8 µm long channel passivated by h-BN. (a) Transfer characteristic measured at 293 K and $V_{ds}$ = 0.1 V before the IL was deposited. (b) $I_{ds}$ as a function of $V_{bg}$ at $V_{ds}$ = 0.1 V at IL gate voltages varying between zero and $V_{ILg}$ = 6 V, obtained after the sample has been cooled down from 230 to 170 K. The inset shows the same data on a linear scale. (c) $I_{ds}$ as a function of $V_{bg}$ at $V_{ds}$ = 0.1 V, measured at $V_{ILg}$=-7 V after the sample has been cooled down from 230 to 170 K. (d-f) Output characteristics of the device at zero applied IL gate voltage, different back-gate and IL gate voltages, measured after the device has been cooled down from 230 K to 170 K with no applied IL gate voltage. The insets in (a), (d), (e) and (f) show schematically the SB height and width at various IL and back gate voltages.

**Figure 3.** (a) Current versus bias voltage characteristics of a graphene/WSe2 Schottky barrier (SB) at fixed back-gate voltages $V_{bg}$ = 20 V and 30 V, showing a Schottky diode behavior. (b) Forward-bias $I_f$-$V_f$ characteristics of a device dominated by a single graphene/WSe$_2$ SB junction measured between 77 K and 160 K. (c) Plot of the saturation current $I_s$ as a function of the inverse characteristic energy $\varphi_0$, obtained from (b). The SB height is determined using the exponential fitting function listed in the figure.

**Figure 4.** Temperature-dependent two-terminal conductivity as a function of the gate voltage, measured after the WSe$_2$ device, characterized in Figure 2, was cooled down from 230 K at (a) $V_{ILg}$ = 6 V and (b)

$V_{ILg}$ = -7 V. (c) Temperature dependent field-effect mobility in the WSe$_2$ device, extracted from two-terminal conductivity measurements as a function of the back-gate voltage at $V_{ILg}$ = 6 V and $V_{ILg}$ = -7 V.

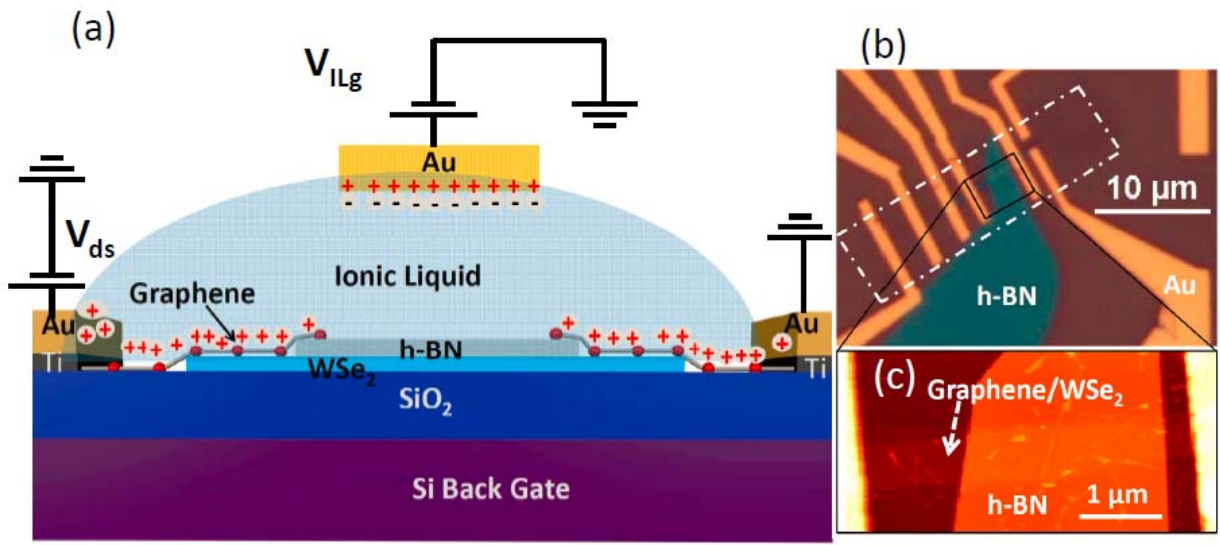

Figure 1

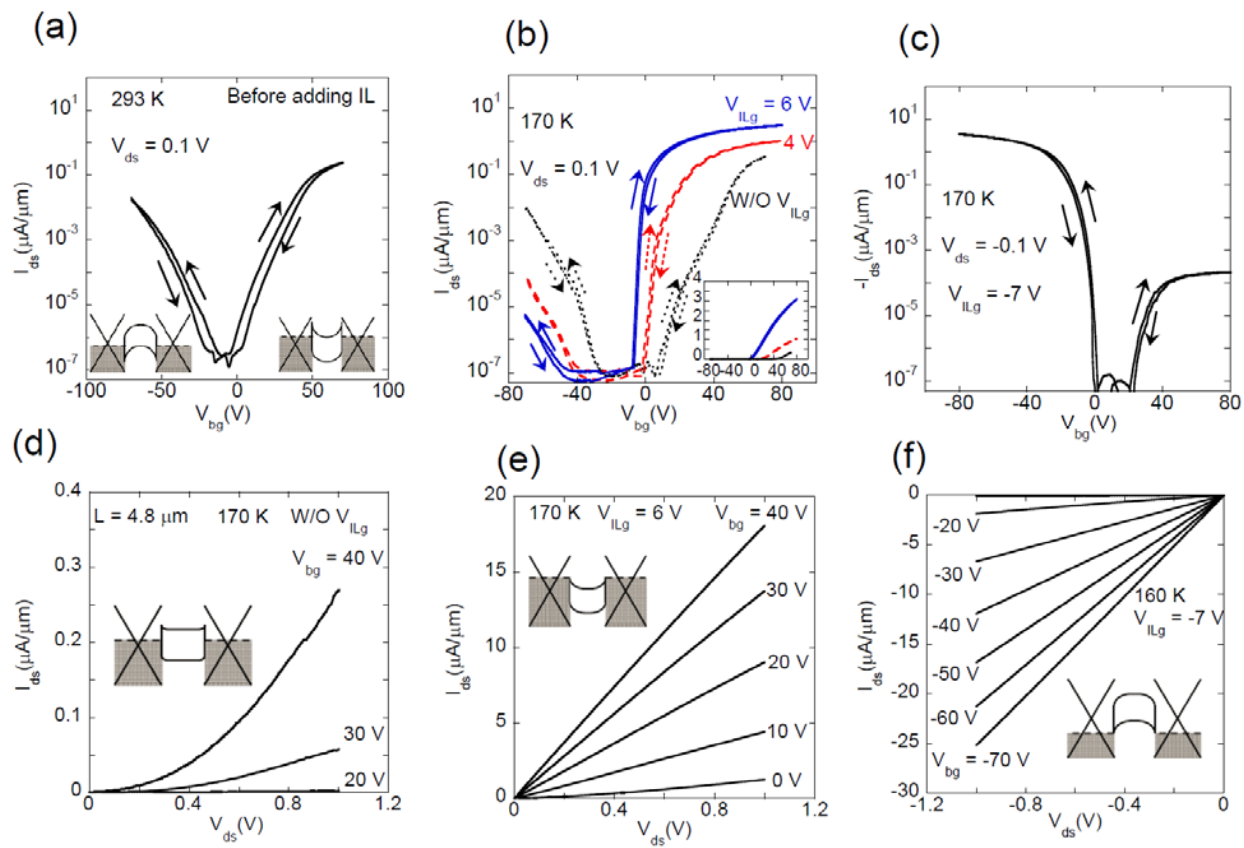

Figure 2

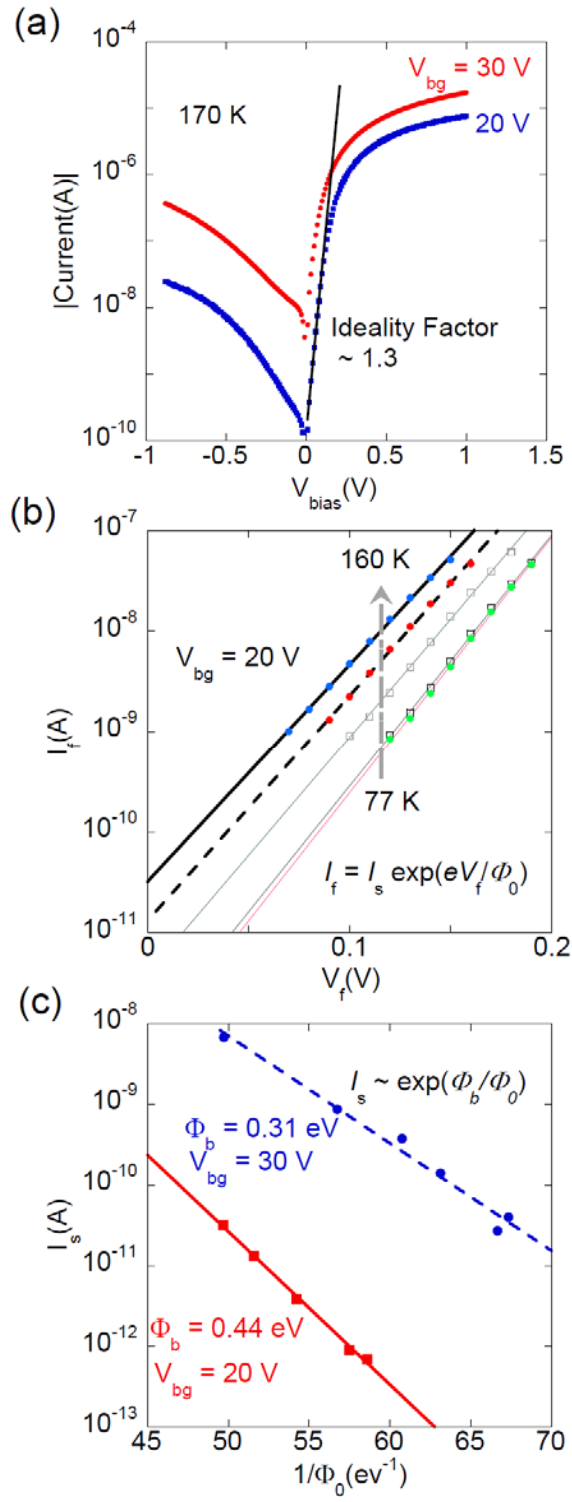

Figure 3

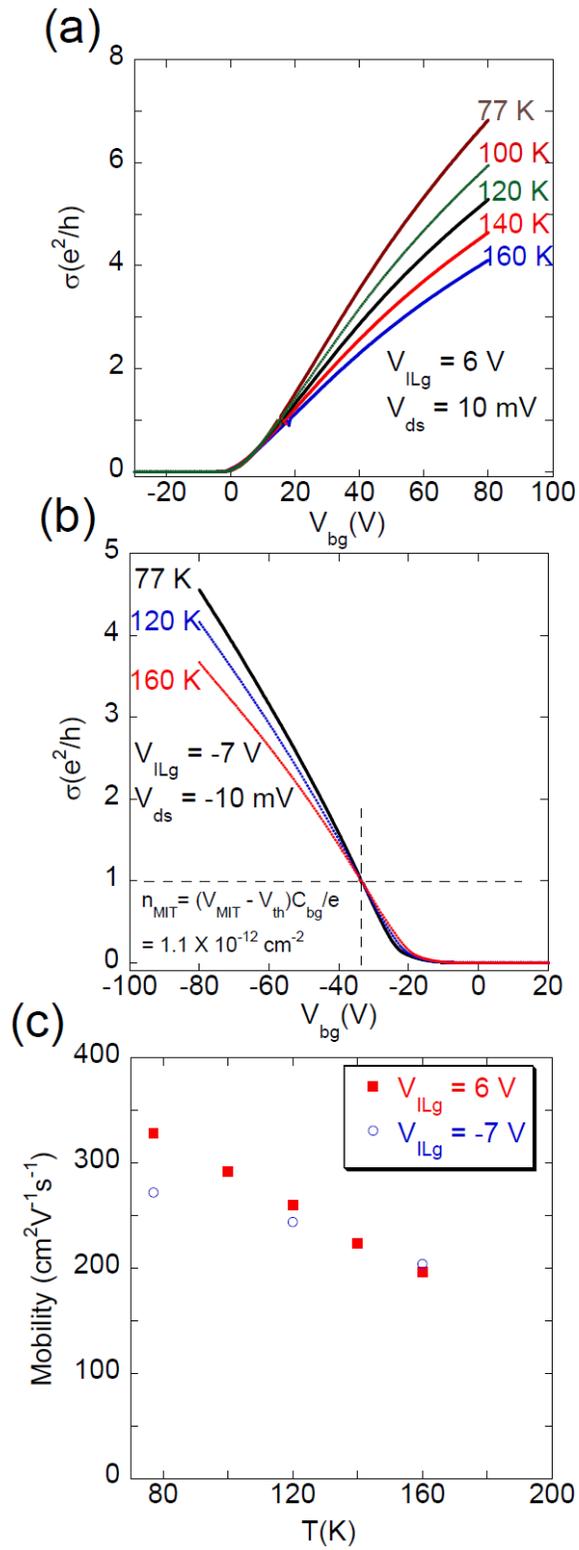

Figure 4

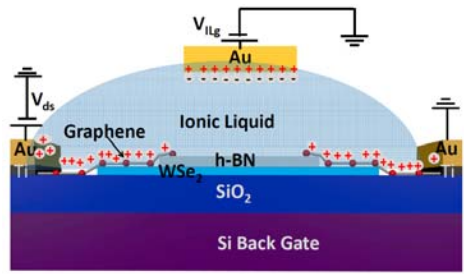

TOC Graphic